\documentclass[aps, amsmath,amssymb,prb,twocolumn,superscriptaddress]{revtex4-2}
\usepackage{graphicx}
\usepackage[colorlinks,bookmarks=false,citecolor=blue,linkcolor=red,urlcolor=blue]{hyperref}
\usepackage[utf8]{inputenc}
\usepackage{amsmath,bm}
\usepackage{bbold}
\usepackage{color,soul}
\usepackage{xstring}
\usepackage{wasysym}
\usepackage{xspace}
\usepackage{amssymb}
\usepackage{time}
\usepackage{bbold}
\usepackage{subfigure}
\usepackage{multirow}
\usepackage{xspace}
\usepackage{url}
\usepackage{amsmath}
\usepackage{booktabs}
\usepackage{lipsum}

\def\<{\langle}
\def\>{\rangle}

\usepackage{ulem}

\begin{document}

\title{
Effective-Geometry Rescaling and Universal Critical Behavior in the Anisotropic Three-State Potts Model on the Square Lattice}
\author{\firstname{Fan} \surname{Yang}}
\affiliation{\mbox{School of Physical Science and Technology, Beijing University of Posts and Telecommunications, Beijing 100876, China}}

\author{\firstname{Jian} \surname{Gao}}
\email{gao.jian@bupt.edu.cn}
\affiliation{\mbox{School of Physical Science and Technology, Beijing University of Posts and Telecommunications, Beijing 100876, China}}

\author{\firstname{Lu} \surname{Liu}}
\email{liulu96@bit.edu.cn}
\affiliation{\mbox{School of Physics, Beijing Institute of Technology, Beijing 100081, China}}

\author{\firstname{Yuhai} \surname{Liu}}
\email{yuhailiu@bupt.edu.cn}
\affiliation{\mbox{School of Physical Science and Technology, Beijing University of Posts and Telecommunications, Beijing 100876, China}}
\date{\today}
\begin{abstract}
We study the two-dimensional anisotropic three-state Potts ferromagnet on the square lattice using Wolff single-cluster Monte Carlo simulations and finite-size scaling. For coupling ratios $\lambda=J_y/J_x=0.5$, $0.75$, and $1$, finite-size scaling of the correlation ratio yields critical
behavior consistent with the two-dimensional three-state Potts universality class. The scaling of the leading Fisher zeros gives a correlation-length exponent consistent with $\nu=5/6$, while their cumulative density is consistent with the expected specific-heat exponent $\alpha=1/3$. We further characterize the anisotropy at criticality using directional correlation ratios $R_x$ and $R_y$ together with directional FK wrapping probabilities. For $\lambda=0.5$ on a physically square lattice, $R_x$ and $R_y$ approach distinct critical values while yielding a common correlation-length exponent. From the wrapping probabilities, we independently determine an effective aspect ratio $\rho_e^\square=0.6413(5)$, in close agreement with the theoretical value $\rho_e^{\square,\mathrm{th}}\simeq0.64150030$ obtained from the isoradial representation. Using the theoretical value to set the physical aspect ratio restores directional equivalence, $R_x\simeq R_y$, and brings the overall correlation ratio toward the isotropic-square reference. The results show that spatial anisotropy changes the effective critical geometry without altering the bulk three-state Potts universality class.
\end{abstract}
\maketitle
\section{Introduction}
Universality is a fundamental organizing principle of critical phenomena. Near a continuous phase transition, the asymptotic critical behavior is governed primarily by spatial dimensionality, order-parameter symmetry, and the range of interactions, rather than by microscopic details \cite{WilsonKogut1974,Fisher1974,Cardy1996}. Consequently, microscopically different systems can share the same critical exponents and thus belong to the same universality class. At the same time, finite-size critical quantities may retain a dependence on system geometry, boundary conditions, and spatial
anisotropy~\cite{KamieniarzBlote1993,Selke2007,Malakis2014}, making their analysis particularly useful for understanding how microscopic anisotropy is encoded in finite-size critical behavior.

The $q$-state Potts model provides a canonical setting in which to examine this distinction \cite{Potts1952,Wu1982,Baxter1973}. On the two-dimensional square lattice, the ferromagnetic model undergoes a continuous transition for $q\leq4$, and the $q=3$ case is particularly well characterized because both the critical point and the critical exponents are known exactly \cite{Wu1982,Baxter1973}. Monte Carlo finite-size scaling has long provided accurate numerical
tests of critical behavior and universality in Potts and related lattice
models \cite{Privman1984,Lee1992,LiuYanWang2021}. When the nearest-neighbor couplings become direction dependent, $J_x\neq J_y$, the permutation symmetry of the Potts states is preserved while spatial rotational symmetry is broken. The anisotropic three-state Potts model thus offers a controlled framework for asking whether spatial anisotropy changes the underlying bulk critical scaling or instead manifests itself primarily through anisotropy-dependent finite-size geometry.

Previous analytical and numerical studies largely support the latter picture. For anisotropic square-lattice Potts ferromagnets, real-space renormalization-group calculations recovered the self-dual critical line and the expected universality flow for $q\leq4$ \cite{deOliveira1982}, while approximate equation-of-state calculations found critical exponents to be insensitive to anisotropy \cite{Chame1988}. Transfer-operator studies of anisotropic Ising and three-state Potts models further showed that apparently anisotropic corner critical behavior can be reconciled with conformal predictions after an appropriate rescaling of lengths, or equivalently through an effective geometric angle \cite{Karevski1997}. More generally, anisotropic critical systems can be described in terms of anisotropy matrices, generalized shear transformations, and multiparameter universality \cite{Dohm2019,DohmWessel2021,Dohm2023}. These results support the view that ordinary ferromagnetic anisotropy modifies the finite-size geometry of the critical system while leaving its universality class unchanged.

This distinction becomes especially important for finite-size dimensionless observables. Critical Binder cumulants and scaling functions may retain explicit information about the spatial anisotropy and the shape of the simulation domain even when the bulk universality class is unchanged \cite{Chen2004,Selke2005}. Hu, Ziff, and Deng showed that anisotropic critical systems can be mapped quantitatively onto isotropic ones through an effective aspect ratio and, for more general geometries, an effective boundary twist. They further formulated the effective geometry of anisotropic $q$-state Potts models using an isoradial mapping and performed numerical tests based on wrapping probabilities on the triangular lattice \cite{HuZiffDeng2022}. These results motivate a more stringent test of
the effective-geometry picture: whether an independently determined
effective geometry can quantitatively account for the directional
splitting of finite-size dimensionless observables in the anisotropic
three-state Potts ferromagnet on the square lattice.

In this work, we combine these ideas to study the anisotropic $q=3$ Potts ferromagnet on the square lattice. Using the Wolff single-cluster algorithm \cite{Wolff1989}, we consider coupling ratios $\lambda=J_y/J_x=0.5$, $0.75$, and $1$. We first use finite-size scaling of the correlation ratio $R$ to locate the transition and determine the correlation-length exponent $\nu$. Such dimensionless correlation ratios provide convenient finite-size probes of continuous phase transitions \cite{Tomita2002}. We also determine the anomalous dimension $\eta$ from the critical scaling of the structure factor. We then analyze Fisher zeros in the complex-temperature plane \cite{Fisher1965,Itzykson1983,KimCreswick1998,Kim2004}, obtaining an independent estimate of $\nu$ from the leading zeros and testing the expected specific-heat exponent through their cumulative density.

To characterize how anisotropy is encoded in the finite-size critical
geometry, we further introduce directional correlation ratios $R_x$ and $R_y$ and analyze directional wrapping probabilities within the Fortuin--Kasteleyn cluster representation \cite{FortuinKasteleyn1972,Pinson1994,ZiffLorenzKleban1999}. The wrapping probabilities provide an independent numerical estimate of the effective aspect ratio $\rho_e$, which is compared with the theoretical prediction from the isoradial construction. The theoretical value is then used, without additional fitting, to construct matched rectangular geometries. This framework allows us to test whether anisotropy-induced directional differences in finite-size observables can be absorbed into an effective geometric rescaling while leaving the bulk universality class unchanged.

The paper is organized as follows. In Sec.~\ref{sec:model and method}, we introduce the model and numerical methods. In Sec.~\ref{sec:results}, we present and discuss the numerical results. Finally, in Sec.~\ref{sec:conclusions}, we summarize our conclusions and discuss the implications of our results.

\section{Model and Methods}
\label{sec:model and method}
\subsection{Anisotropic three-state Potts model}
\label{sec:model}
We consider the ferromagnetic three-state Potts model on a square lattice with periodic boundary conditions. The Hamiltonian is
\begin{equation}
\mathcal{H}
=-J_x\sum_{\langle ij\rangle_x}\delta_{\sigma_i,\sigma_j}
-J_y\sum_{\langle ij\rangle_y}\delta_{\sigma_i,\sigma_j},
\label{eq:H}
\end{equation}
where $\sigma_i\in\{0,1,2\}$, and $\langle ij\rangle_x$ and $\langle ij\rangle_y$ denote nearest-neighbor pairs along the horizontal and vertical directions, respectively, with each bond counted once. The couplings are ferromagnetic, $J_x,J_y>0$. We set $k_{\rm B}=1$ and $J_x=1$, and define the anisotropy ratio $\lambda=\frac{J_y}{J_x}$ with $\lambda=1$ corresponding to the isotropic model. In this work, we consider $\lambda=0.5$, $0.75$, and $1$.

\begin{figure}[t]
	\centering
	\includegraphics[width=\linewidth]{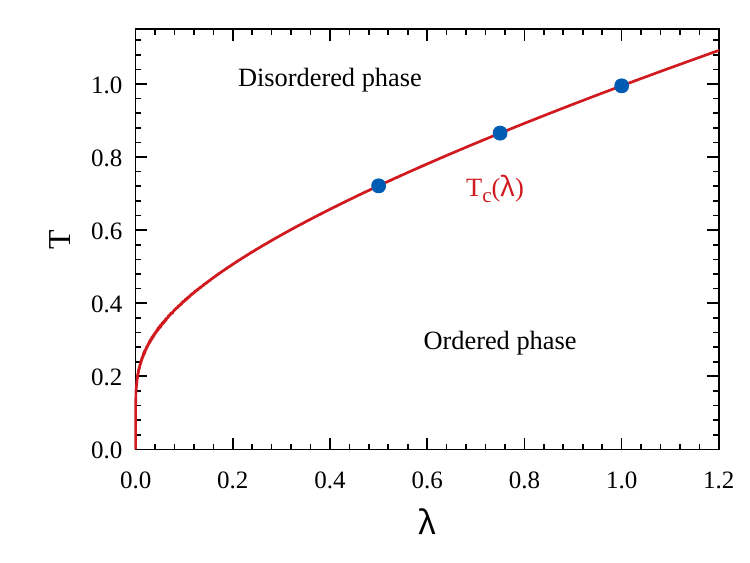}
	\caption{Phase diagram of the anisotropic two-dimensional ferromagnetic three-state Potts model in the $\lambda$--$T$ plane. The red solid line denotes the exact critical line determined by the self-dual condition, and the blue circles mark the coupling ratios $\lambda=0.5$, $0.75$, and $1$ studied in this work. The regions below and above the critical line correspond to the ordered and disordered phases, respectively.}
\label{fig1}
\end{figure}

For the anisotropic square-lattice Potts model, the critical temperature is exactly determined by the self-dual condition
\cite{Wu1982},
\begin{equation}
\left(e^{\beta_c J_x}-1\right)
\left(e^{\beta_c J_y}-1\right)=q,
\label{eq:critical-line}
\end{equation}
where $\beta_c=1/T_c$ and $q=3$ throughout this work. As $\lambda$ decreases, the coupling in the $y$ direction weakens and the critical temperature decreases. In the limit $\lambda\rightarrow0$, the system approaches a set of decoupled one-dimensional Potts chains and $T_c$ vanishes \cite{Wu1982}. The corresponding phase diagram is shown in Figure~\ref{fig1}.

\subsection{Monte Carlo simulations and correlation ratios}
\label{sec:mc}
Monte Carlo sampling is performed using the Wolff single-cluster algorithm~\cite{Wolff1989}. A randomly selected site serves as the seed, and nearest-neighbor spins in the same state as the cluster spins are added with direction-dependent bond-activation probabilities
\begin{equation}
p_x=1-e^{-\beta J_x},
\qquad
p_y=1-e^{-\beta J_y}.
\label{eq:wolff-probability}
\end{equation}

We represent each Potts spin as $z_j=\exp(i2\pi\sigma_j/3)$ and define the structure factor as
\begin{equation}
S(\bm{k})
=\frac{1}{N}
\left\langle
\left|
\sum_{j=1}^{N}
z_j e^{i\bm{k}\cdot\bm{r}_j}
\right|^2
\right\rangle,
\label{eq:Sk}
\end{equation}
where $N=L_xL_y$. For a system of size $L_x\times L_y$, the smallest nonzero wave vectors along the two lattice directions are $\bm{k}_x=\left(2\pi/L_x,0\right)$ and $\bm{k}_y=\left(0,2\pi/L_y\right)$.
We define the corresponding symmetrized structure factors as $S_x=[S(\bm{k}_x)+S(-\bm{k}_x)]/2$ and
$S_y=[S(\bm{k}_y)+S(-\bm{k}_y)]/2$. The directional correlation ratios are then defined as
\begin{equation}
R_x=1-\frac{S_x}{S(\mathbf{0})},
\qquad
R_y=1-\frac{S_y}{S(\mathbf{0})}.
\label{eq:directional-R}
\end{equation}
The overall correlation ratio is defined as
\begin{equation}
R=\frac{R_x+R_y}{2}=1-\frac{S_x+S_y}{2S(\mathbf{0})}.
\label{eq:correlation-ratio}
\end{equation}
In all cases, the structure factors are ensemble averaged before the ratios are evaluated.

For the square systems used in the bulk finite-size analysis, $L_x=L_y=L$, the finite-size scaling form near criticality is
\begin{equation}
R(T,L)=f_R\left((T-T_c)L^{1/\nu},L^{-\omega}\right),
\label{eq:R-FSS}
\end{equation}
where $\nu$ is the correlation-length critical exponent and $\omega$ is the leading correction-to-scaling exponent. The same scaling form is used for $R_x$ and $R_y$ when their directional behavior is analyzed.

\subsection{Fisher zeros and finite-size scaling}
\label{sec:fisher-method}

As an independent probe of the critical behavior, we analyze Fisher zeros of the partition function in the complex inverse-temperature plane. For fixed $J_x$ and $J_y$, Monte Carlo sampling is performed at the exact critical inverse temperature $\beta_0=\beta_c^{\rm sd}$.
From the normalized energy histogram $H_{\beta_0}(E)$, histogram reweighting gives \cite{FerrenbergSwendsen1988}
\begin{equation}
\mathcal{Z}(\Delta\beta)
=
\sum_E H_{\beta_0}(E)
e^{-\Delta\beta(E-E_{\rm ref})}.
\label{eq:reweight}
\end{equation}
Here, $\mathcal{Z}(\Delta\beta)$ is the normalized reweighted partition-function estimator, and $E_{\rm ref}$ is a reference energy introduced to improve numerical stability. The Fisher zeros are defined by the complex roots $\Delta\beta_j$ of $\mathcal{Z}(\Delta\beta)$, where $j=1,2,\ldots$ labels the zeros in order of increasing $|\operatorname{Im}\beta_j|$, with $j=1$ denoting the leading zero. The corresponding zeros in the complex
inverse-temperature plane are $\beta_j=\beta_0+\Delta\beta_j$, where $\Delta\beta$ denotes the complex inverse-temperature displacement from the sampling point $\beta_0$. In the infinite-sampling limit, $\mathcal{Z}$ differs from the partition function only by a nonzero multiplicative factor and therefore has the same zeros.

For a continuous transition, the imaginary part of the leading zero obeys \cite{JankeKenna2002}
\begin{equation}
\operatorname{Im}\beta_1(L)
\propto L^{-1/\nu},
\label{eq:fisher-scaling}
\end{equation}
providing an estimate of $\nu$ independent of the correlation-ratio analysis. The specific-heat exponent is estimated from the cumulative density of zeros~\cite{JankeKenna2002,Sarkanych2026},
\begin{equation}
G_L(r_j)=\frac{2j-1}{2L^2},
\qquad
r_j=|\operatorname{Im}\beta_j|,
\label{eq:zero-density}
\end{equation}
which behaves in the thermodynamic limit as
\begin{equation}
G(r)\propto r^{2-\alpha}.
\label{eq:zero-density-scaling}
\end{equation}

\subsection{Fortuin--Kasteleyn clusters and wrapping probabilities}
\label{sec:wrapping-method}
To characterize the finite-size geometric effects of anisotropy, we measure directional wrapping probabilities in the Fortuin--Kasteleyn (FK) cluster representation \cite{FortuinKasteleyn1972}. For each equilibrium spin configuration, bonds between nearest-neighbor spins in the same Potts state are activated with the direction-dependent probabilities given in Eq.~\eqref{eq:wolff-probability}. The resulting FK clusters are then examined for nontrivial winding around the periodic boundaries. For each FK configuration, we define $I_\mu$ ($\mu=x,y$) as the indicator that at least one FK cluster wraps around the periodic boundary in the $\mu$ direction. The corresponding directional wrapping probabilities are
$P_x=\langle I_x\rangle$ and $P_y=\langle I_y\rangle$. A configuration wrapping in both directions contributes to both
$P_x$ and $P_y$.

At criticality, an anisotropic system can be associated with an isotropic rectangular system characterized by an effective aspect ratio $\rho_e$ \cite{HuZiffDeng2022}. For a physically square anisotropic system, we denote the corresponding effective aspect ratio by $\rho_e^\square$. We determine $\rho_e^\square$ by matching the directional wrapping probabilities,
\begin{equation}
P_\mu^{\rm aniso}(\lambda,L)
\simeq
P_\mu^{\rm iso}(\rho_e^\square),
\qquad \mu=x,y,
\label{eq:rho-wrapping}
\end{equation}
where $P_\mu^{\rm iso}(\rho)$ denotes the corresponding critical wrapping probability of the isotropic model at aspect ratio $\rho$, with the convention $\rho=L_x/L_y$. The same wrapping-event definition and periodic boundary conditions are used for both systems. Finite-size estimates of $\rho_e^\square$ are extrapolated to the thermodynamic limit.

\section{Numerical Results}
\label{sec:results}
\subsection{Bulk critical scaling and universality}
\label{sec:bulk-scaling}

\begin{figure}[htbp]
\centering
\includegraphics[width=\linewidth]{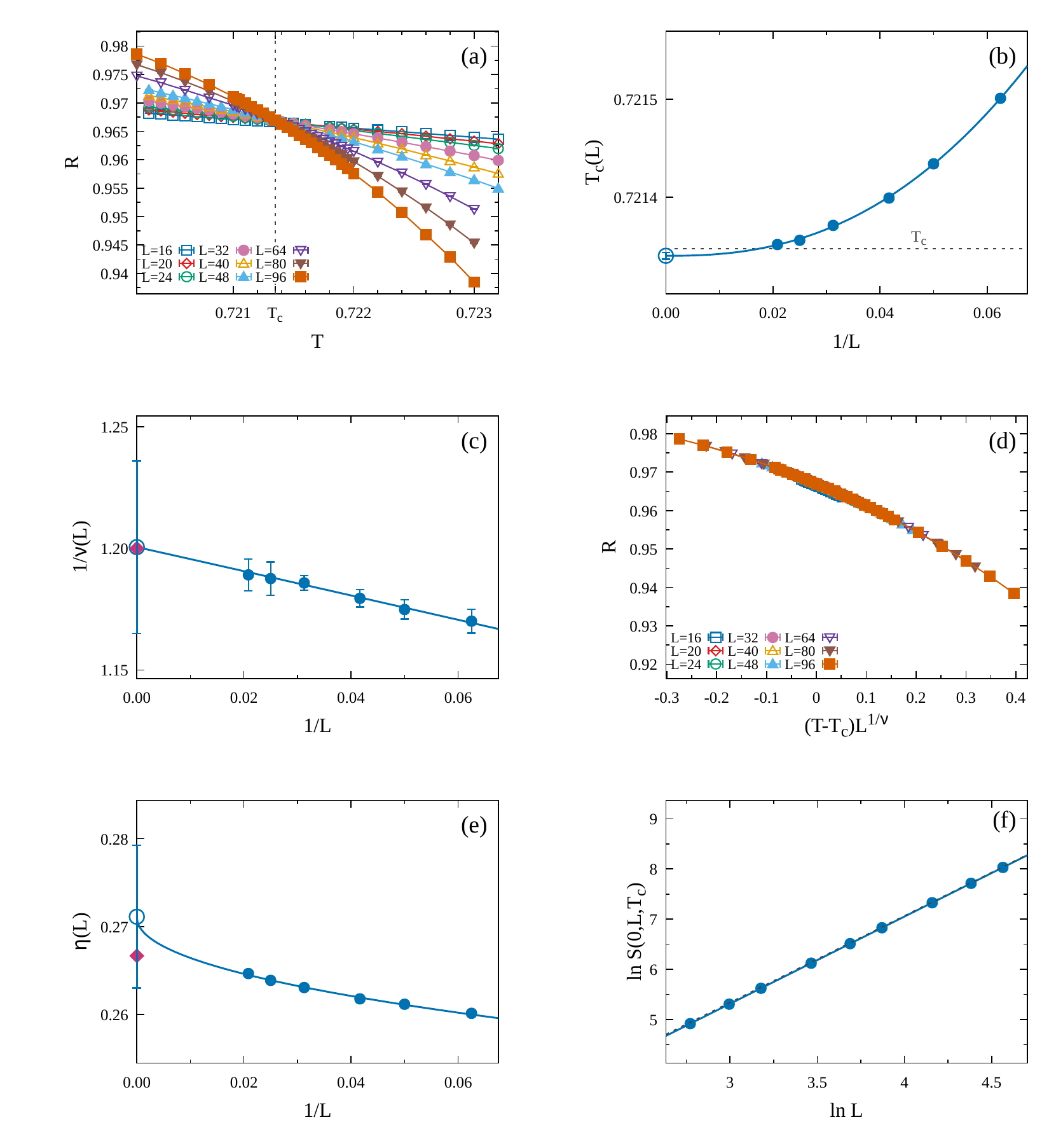}
\caption{Finite-size scaling analysis at $\lambda=0.5$. (a) Correlation ratio $R$ for different system sizes. (b) Crossing temperatures $T_c(L)$, extrapolating to $T_c=0.721340(3)$. (c) Effective $1/\nu(L)$ obtained from the slope ratios, with extrapolated value $1/\nu=1.20(3)$. (d) Data collapse of $R$ using $T_c^{\rm sd}\simeq0.7213475$ and $1/\nu=6/5$. (e) Effective $\eta(L)$, extrapolating to $\eta=0.271(8)$. (f) Log--log scaling of $S(\mathbf{0},L,T_c^{\rm sd})$ with the exact exponent $\eta=4/15$. Open symbols in (b), (c), and (e) denote the thermodynamic-limit estimates; dashed lines in (a) and (b) indicate $T_c^{\rm sd}$, while colored markers on the vertical axes in (c) and (e) indicate the exact values $1/\nu=6/5$ and $\eta=4/15$, respectively.}
\label{fig:corr-main}
\end{figure}

We first examine the bulk critical behavior from the finite-size
scaling of the overall correlation ratio defined in Eq.~\eqref{eq:correlation-ratio}. Figure~\ref{fig:corr-main} shows the results for $\lambda=0.5$. As shown in Figure~\ref{fig:corr-main}(a), the curves of $R(T,L)$ for different system sizes exhibit a clear crossing near the critical temperature. We define the finite-size crossing temperature $T_c(L)$ for the pair $(L,2L)$ by
$R[T_c(L),L]=R[T_c(L),2L]$. The resulting $T_c(L)$ values are shown in Figure~\ref{fig:corr-main}(b).
Their finite-size dependence is extrapolated using
\begin{equation}
O(L)=O+a_1L^{-b_1},
\label{eq:O-extrapolation}
\end{equation}
where $O$ denotes the thermodynamic-limit value of the quantity under consideration, $a_1$ is a nonuniversal amplitude, and $b_1$ is an effective finite-size correction exponent. For the crossing temperatures, we take $O(L)=T_c(L)$ and $O=T_c$. The extrapolated value $T_c=0.721340(3)$ is very close to the exact self-dual result $T_c^{\rm sd}\simeq0.7213475$ obtained from Eq.~\eqref{eq:critical-line}.

The correlation-length exponent is estimated from the ratio of the slopes of the two correlation-ratio curves at their crossing. For each pair $(L,2L)$, we define
\begin{equation}
\frac{1}{\nu(L)}
=\frac{1}{\ln 2}\ln\left[\frac{\left.\dfrac{\partial R(T,2L)}{\partial T}\right|_{T=T_c(L)}}
{\left.\dfrac{\partial R(T,L)}{\partial T}\right|_{T=T_c(L)}}\right].
\label{eq:nu-effective}
\end{equation}
The resulting estimates of $1/\nu(L)$ are shown in Figure~\ref{fig:corr-main}(c). Their thermodynamic-limit extrapolation according to Eq.~\eqref{eq:O-extrapolation} with $O=1/\nu$ gives $1/\nu=1.20(3)$, consistent with the exact three-state Potts value $1/\nu=6/5$. We further examine the finite-size scaling of the dimensionless correlation ratio through data collapse. Near the critical temperature, its scaling form satisfies Eq.~\eqref{eq:R-FSS}. In Figure~\ref{fig:corr-main}(d), we plot $R(T,L)$ as a function of $(T-T_c^{\rm sd})L^{6/5}$ using the exact self-dual critical temperature $T_c^{\rm sd}$ and the exact three-state Potts value $1/\nu=6/5$. The data for different system sizes collapse onto a common curve in
the critical region on both sides of the crossing point, providing further support for the three-state Potts value of the correlation-length exponent.

The anomalous dimension $\eta$ is further estimated from the finite-size scaling of the zero-momentum structure factor. For the present two-dimensional classical system, its scaling form can be written as
\begin{equation}
S(\mathbf{0},L,T)=L^{2-\eta}f_S\left[(T-T_c)L^{1/\nu},L^{-\omega}\right].
\label{eq:S-FSS}
\end{equation}
At the critical temperature, the effective exponent for a pair of system sizes $(L,rL)$ is therefore given by
\begin{equation}
\eta(L)=2-\frac{1}{\ln r}\ln\left[\frac{S(\mathbf{0},rL,T_c)}{S(\mathbf{0},L,T_c)}\right].
\label{eq:eta-effective}
\end{equation}
In our analysis, we take $r=2$ and evaluate the structure factors directly at the exact critical temperature $T_c^{\rm sd}$. The resulting $\eta(L)$ values are shown in Figure~\ref{fig:corr-main}(e). Their thermodynamic-limit extrapolation according to Eq.~\eqref{eq:O-extrapolation} with $O=\eta$ gives $\eta=0.271(8)$, consistent with the exact three-state Potts value $\eta=4/15$. As a consistency check, the leading finite-size behavior of the structure factor at the critical temperature is
$\ln S(\mathbf{0},L,T_c)=(2-\eta)\ln L+\mathrm{const.}+\cdots$. Figure~\ref{fig:corr-main}(f) shows $S(\mathbf{0},L,T_c^{\rm sd})$ on logarithmic axes. The data closely follow the reference behavior $S_{\rm ref}(L)=CL^{26/15}$ corresponding to the exact three-state Potts value $\eta=4/15$, where $C$ is chosen only for visual comparison.

The close agreement of the extrapolated critical temperature with the exact self-dual value supports the location of the transition, while the consistency of $\nu$ and $\eta$ with the exact three-state Potts
exponents provides evidence for three-state Potts universality in the bulk critical behavior at $\lambda=0.5$.

\begin{figure}[htbp]
\centering
\includegraphics[width=\linewidth]{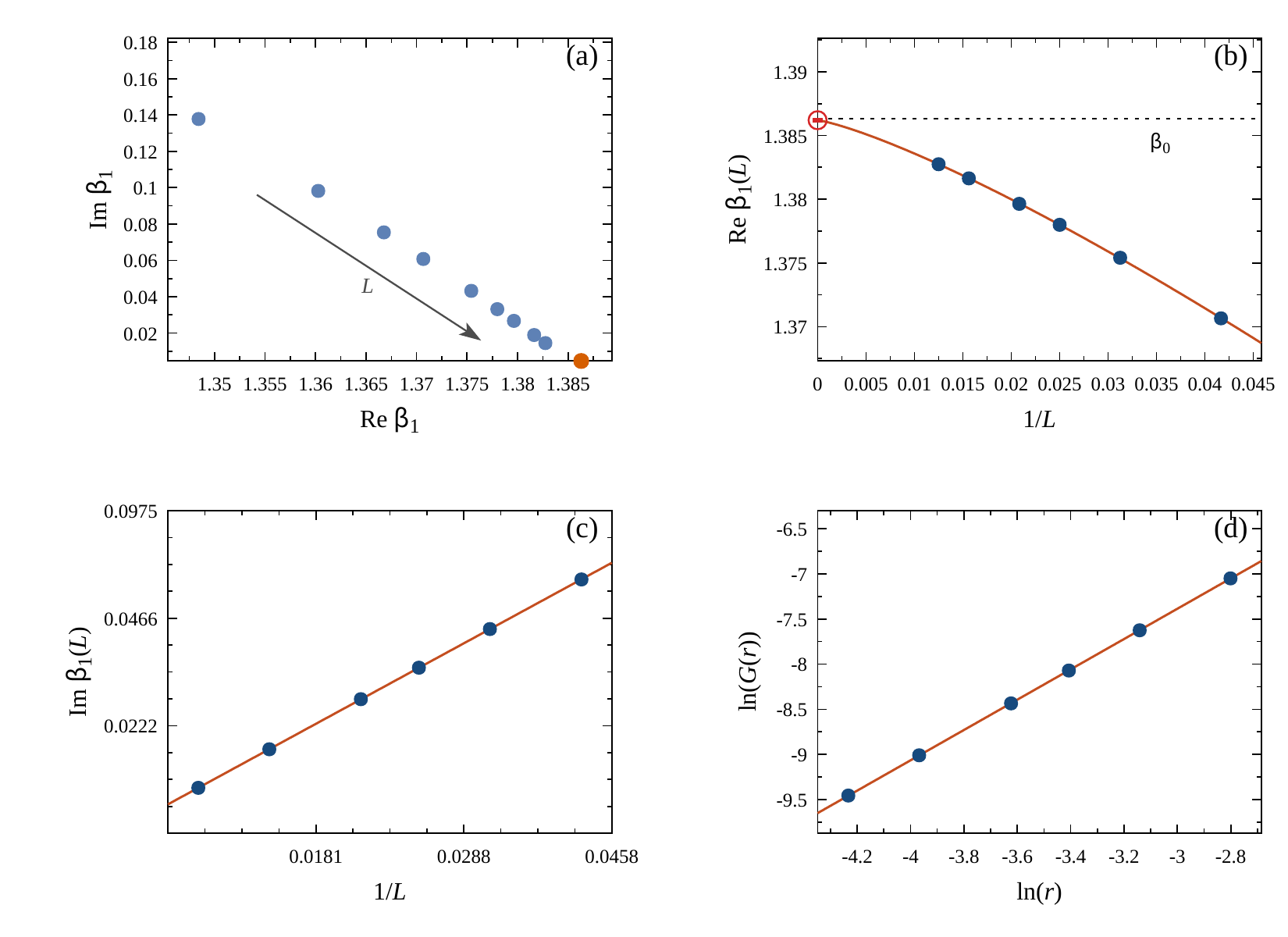}
\caption{Analysis of the leading Fisher zero at $\lambda=0.5$. (a) Leading Fisher zeros $\beta_1(L)$ in the upper half of the complex inverse-temperature plane for different system sizes $L$. The
horizontal and vertical axes represent $\operatorname{Re}\beta_1(L)$ and $\operatorname{Im}\beta_1(L)$, respectively, and the arrow indicates the direction of increasing $L$. (b) Real part of the leading zero as a function of $1/L$, together with its extrapolation to the thermodynamic limit. The dashed
horizontal line indicates the exact self-dual critical inverse temperature, $\beta_c^{\rm sd}=\ln 4$.
(c) Imaginary part of the leading zero as a function of $1/L$ on logarithmic axes, analyzed using Eq.~\eqref{eq:fisher-scaling}. (d) Cumulative density $G_L(r_1)=1/(2L^2)$ as a function of $r_1=|\operatorname{Im}\beta_1|$ on logarithmic axes. The fitted slope yields $2-\alpha$. Solid lines denote fits. The red filled point in panel (a) represents the sampling point $\beta_0=\beta_c^{\rm sd}$, while the open symbol in panel (b) denotes the thermodynamic-limit extrapolation.}
\label{fig:fisher}
\end{figure}

We next test this conclusion independently using Fisher zeros in the complex inverse-temperature plane. Specifically, we examine the finite-size behavior of the Fisher zeros obtained from Eq.~\eqref{eq:reweight}, using the exact self-dual critical inverse temperature $\beta_0=\beta_c^{\rm sd}=\ln 4$ as the sampling point.
Figure~\ref{fig:fisher}(a) shows the leading Fisher zero $\beta_1(L)$ in the upper half of the complex inverse-temperature plane.  As $L$ increases, $\operatorname{Im}\beta_1(L)$ decreases while $\operatorname{Re}\beta_1(L)$ approaches $\beta_c^{\rm sd}$, consistent with the accumulation of Fisher zeros at the critical point in the thermodynamic limit.
To quantify this convergence, we extrapolate the real part of the
leading zero using
\begin{equation}
\operatorname{Re}\beta_1(L)=\beta_c+aL^{-p},
\label{eq:real-zero-extrapolation}
\end{equation}
where $a$ is a nonuniversal amplitude and $p$ is an effective finite-size correction exponent. The extrapolation in Figure~\ref{fig:fisher}(b) yields $\beta_c=1.38632(4)$, in excellent agreement with the exact self-dual value
$\beta_c^{\rm sd}=\ln 4$. Although the histogram is sampled at $\beta_0=\beta_c^{\rm sd}$, this choice does not constrain the
extrapolated real part of the leading zero to the same value.
The agreement therefore provides an independent check of the
critical-point location.

We next examine the imaginary part of the leading Fisher zero. Figure~\ref{fig:fisher}(c) shows $\operatorname{Im}\beta_1(L)$ as a
function of $1/L$ on logarithmic axes. According to Eq.~\eqref{eq:fisher-scaling}, the asymptotic slope gives $1/\nu$.
A power-law fit yields $1/\nu_{\rm zero}=1.198(2)$, in good agreement with the exact three-state Potts value $1/\nu=6/5$.

For the cumulative-density analysis, Figure~\ref{fig:fisher}(d) shows $G_L(r_1)=1/(2L^2)$ as a function of
$r_1=|\operatorname{Im}\beta_1|$ on logarithmic axes. According to Eq.~\eqref{eq:zero-density-scaling}, the slope gives $2-\alpha$. The fit yields $\alpha=0.323(8)$, consistent with the exact three-state Potts value $\alpha=1/3$.

Together with $G_L(r_1)\propto L^{-2}$, the scaling forms in Eqs.~\eqref{eq:fisher-scaling} and
\eqref{eq:zero-density-scaling} imply the two-dimensional hyperscaling relation $2-\alpha=2\nu$. The exponent estimates obtained from the leading-zero scaling and cumulative-density analysis are
consistent with this relation within the present uncertainties.

Overall, the Fisher-zero analysis independently locates the critical point and yields critical exponents consistent with the exact three-state Potts values.

Table~\ref{tab:critical-results} summarizes the critical parameters obtained from finite-size scaling and Fisher-zero analyses for $\lambda=0.5$, $0.75$, and $1$. The two approaches provide complementary estimates of the bulk critical behavior.

\begin{table*}[t]
\centering
\caption{Critical parameters obtained from real-temperature finite-size scaling and Fisher-zero analyses for different anisotropy ratios $\lambda$. The estimates are based on the correlation ratio, structure factor, leading Fisher zero, and cumulative zero density. $T_c^{\rm sd}$ denotes the exact self-dual critical temperature, and the final row gives the exact three-state Potts critical exponents.}
\label{tab:critical-results}
\small
\setlength{\tabcolsep}{3.5pt}
\renewcommand{\arraystretch}{1.25}
\begin{tabular}{ccccccc}
\toprule
$\lambda$
& \shortstack{$T_c$\\from $R$}
& $T_c^{\rm sd}$
& \shortstack{$1/\nu$\\from $R$}
& $\eta$
& \shortstack{$1/\nu$\\from Fisher zeros}
& \shortstack{$\alpha$\\from $G_L$} \\
\midrule
$0.50$
& $0.721340(3)$
& $0.7213475$
& $1.20(3)$
& $0.271(8)$
& $1.198(2)$
& $0.323(8)$ \\
$0.75$
& $0.865413(2)$
& $0.8654155$
& $1.193(5)$
& $0.2630(3)$
& $1.204(3)$
& $0.324(7)$ \\
$1.00$
& $0.994979(5)$
& $0.9949729$
& $1.198(2)$
& $0.265(1)$
& $1.199(3)$
& $0.325(4)$ \\
Exact exponents
& ---
& ---
& $6/5$
& $4/15$
& $6/5$
& $1/3$ \\
\bottomrule
\end{tabular}
\end{table*}

Across the three anisotropy ratios, the critical exponents obtained from finite-size scaling and Fisher-zero analysis remain close to the exact three-state Potts values. In particular, the Fisher-zero estimates of $1/\nu$ and $\alpha$ are consistent with the exact values $6/5$ and $1/3$, respectively, within the quoted uncertainties. Together with the estimates of $1/\nu$ and $\eta$ from the correlation-ratio and structure-factor analyses, these results show
that spatial anisotropy changes the nonuniversal critical temperature while leaving the bulk critical behavior consistent with the two-dimensional three-state Potts universality class.

\subsection{Directional critical behavior and effective geometry}
\label{sec:directional-critical}

Having established the bulk critical behavior, we next examine how spatial anisotropy manifests itself in direction-resolved observables. For $\lambda\neq1$, the couplings along the two lattice directions are inequivalent, and the corresponding correlation ratios $R_x$ and $R_y$ exhibit distinct finite-size behavior. A natural question is whether these directional differences persist in the thermodynamic limit or instead represent anisotropic finite-size effects associated with a common bulk critical point.

\begin{figure}[htbp]
\centering
\includegraphics[width=\linewidth]{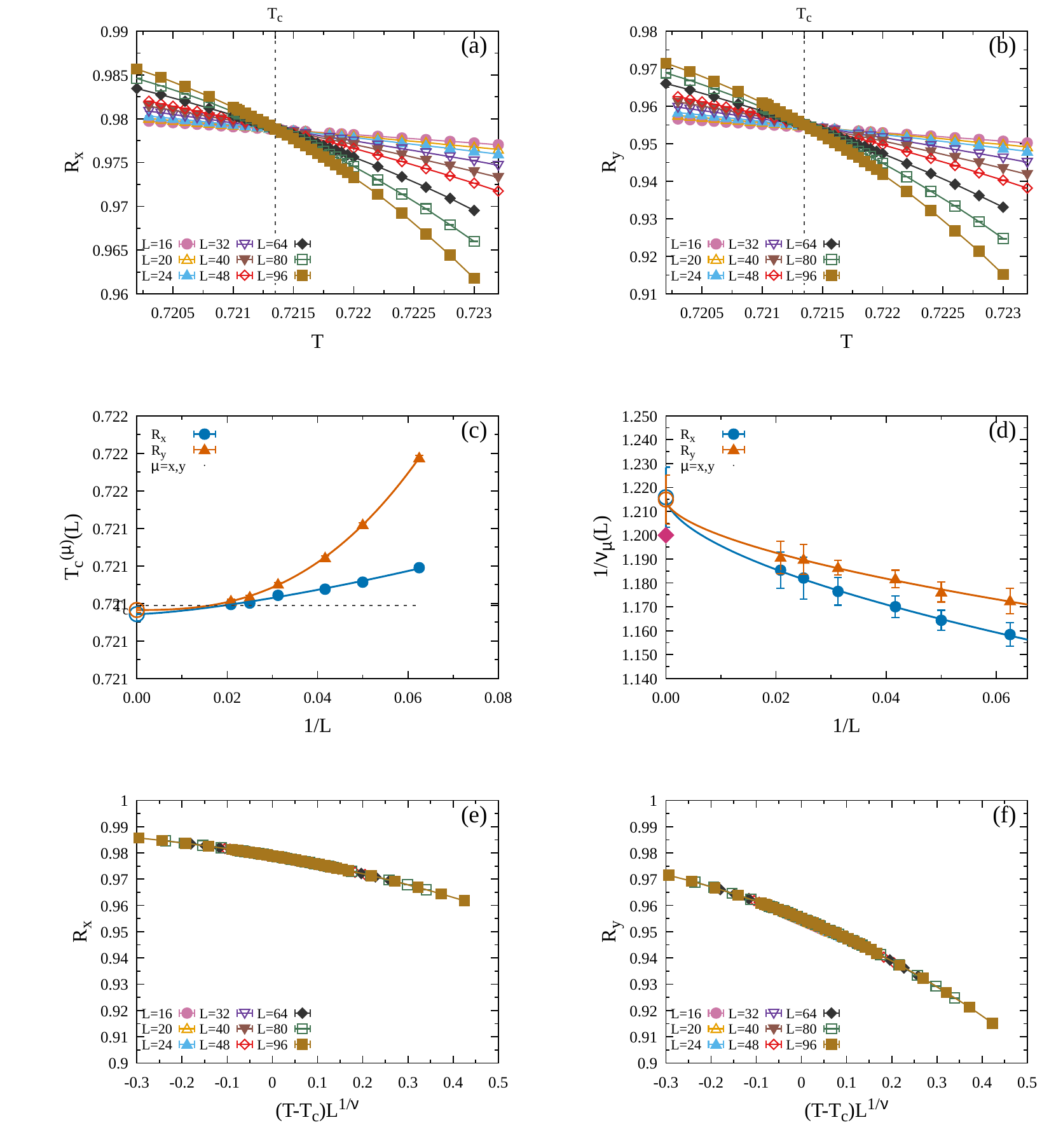}
\caption{Directional finite-size scaling analysis at $\lambda=0.5$.
(a), (b) Directional correlation ratios $R_x$ and $R_y$ as functions
of temperature for different system sizes.
(c) Directional crossing temperatures $T_c^{(\mu)}(L)$,
$\mu=x,y$, for size pairs $(L,2L)$, plotted as functions of $1/L$.
The extrapolations yield
$T_c^{(x)}=0.721336(9)$ and
$T_c^{(y)}=0.721341(3)$ in the thermodynamic limit.
(d) Effective inverse correlation-length exponent estimates
$1/\nu_\mu(L)$, $\mu=x,y$, obtained from the directional slope
ratios and plotted against $1/L$. The extrapolations using
Eq.~\eqref{eq:O-extrapolation} give
$(1/\nu)_{x,\infty}=1.21(1)$ and
$(1/\nu)_{y,\infty}=1.21(1)$.
(e), (f) Data collapse of $R_x$ and $R_y$, respectively, plotted
against $(T-T_c^{\rm sd})L^{6/5}$.
Solid curves in (c) and (d) show the fits, and open circles at
$1/L=0$ denote the extrapolated limits. Dashed lines in (a)--(c)
indicate the self-dual critical temperature. The red diamond in
(d) marks the exact inverse correlation-length exponent
$1/\nu=6/5$.}
\label{fig:directional}
\end{figure}

Figure~\ref{fig:directional}(a) and (b) show the directional correlation ratios $R_x$ and $R_y$, respectively, for $\lambda=0.5$. Both quantities exhibit well-defined crossings near the exact self-dual critical temperature, but their finite-size evolution is visibly different. In particular, the crossing drift is considerably stronger for $R_y$ than for $R_x$, indicating pronounced directional finite-size effects.

For each direction $\mu=x,y$, we define the crossing temperature $T_c^{(\mu)}(L)$ for the size pair $(L,2L)$ by $R_\mu[T_c^{(\mu)}(L),L]=R_\mu[T_c^{(\mu)}(L),2L]$. The resulting crossing temperatures are shown in Figure~\ref{fig:directional}(c). Although $T_c^{(x)}(L)$ and $T_c^{(y)}(L)$ follow clearly different finite-size trajectories, both approach the same critical temperature as $L$ increases. Extrapolation to the thermodynamic limit gives $T_c^{(x)}=0.721336(9)$ and $T_c^{(y)}=0.721341(3)$, both close to the exact self-dual value $T_c^{\rm sd}\simeq0.7213475$. Thus, the directional correlation ratios do not indicate distinct transition temperatures; rather, their finite-size crossing trajectories converge toward a common critical point.

We next determine the correlation-length critical exponent $\nu$ separately from the two directional correlation ratios. In analogy with Eq.~\eqref{eq:nu-effective}, the directional effective inverse exponents $1/\nu_\mu(L)$ are obtained by replacing $R$ with $R_\mu$
and evaluating the slope ratio at the corresponding crossing temperature $T_c^{(\mu)}(L)$, with $\mu=x,y$. As shown in Figure~\ref{fig:directional}(d), the resulting $1/\nu_\mu(L)$ estimates exhibit different finite-size dependences in the two directions. Nevertheless, their thermodynamic-limit extrapolations, $(1/\nu)_{x,\infty}=1.21(1)$ and $(1/\nu)_{y,\infty}=1.21(1)$, are mutually consistent and agree with the exact three-state Potts value $1/\nu=6/5$ within the quoted uncertainties. The directional dependence observed at finite size therefore does not imply distinct correlation-length exponents in the
two directions.

This conclusion is further supported by the scaling collapses shown in Figure~\ref{fig:directional}(e) and (f). Using the exact self-dual critical temperature $T_c^{\rm sd}$ and the exact thermal exponent $1/\nu=6/5$, the data for different system sizes collapse onto
well-defined scaling curves for both $R_x$ and $R_y$. The two scaling functions need not coincide, since their detailed forms may contain nonuniversal information associated with the anisotropic spatial metric. Nevertheless, the successful collapses using the same critical
temperature and correlation-length exponent further support a common
asymptotic thermal critical scaling in the two directions.

Taken together, these results show that the directional finite-size differences do not correspond to distinct thermal critical behavior. We therefore next examine how the remaining anisotropy is encoded in the critical finite-size geometry.

\begin{figure}[htbp]
\centering
\includegraphics[width=\linewidth]{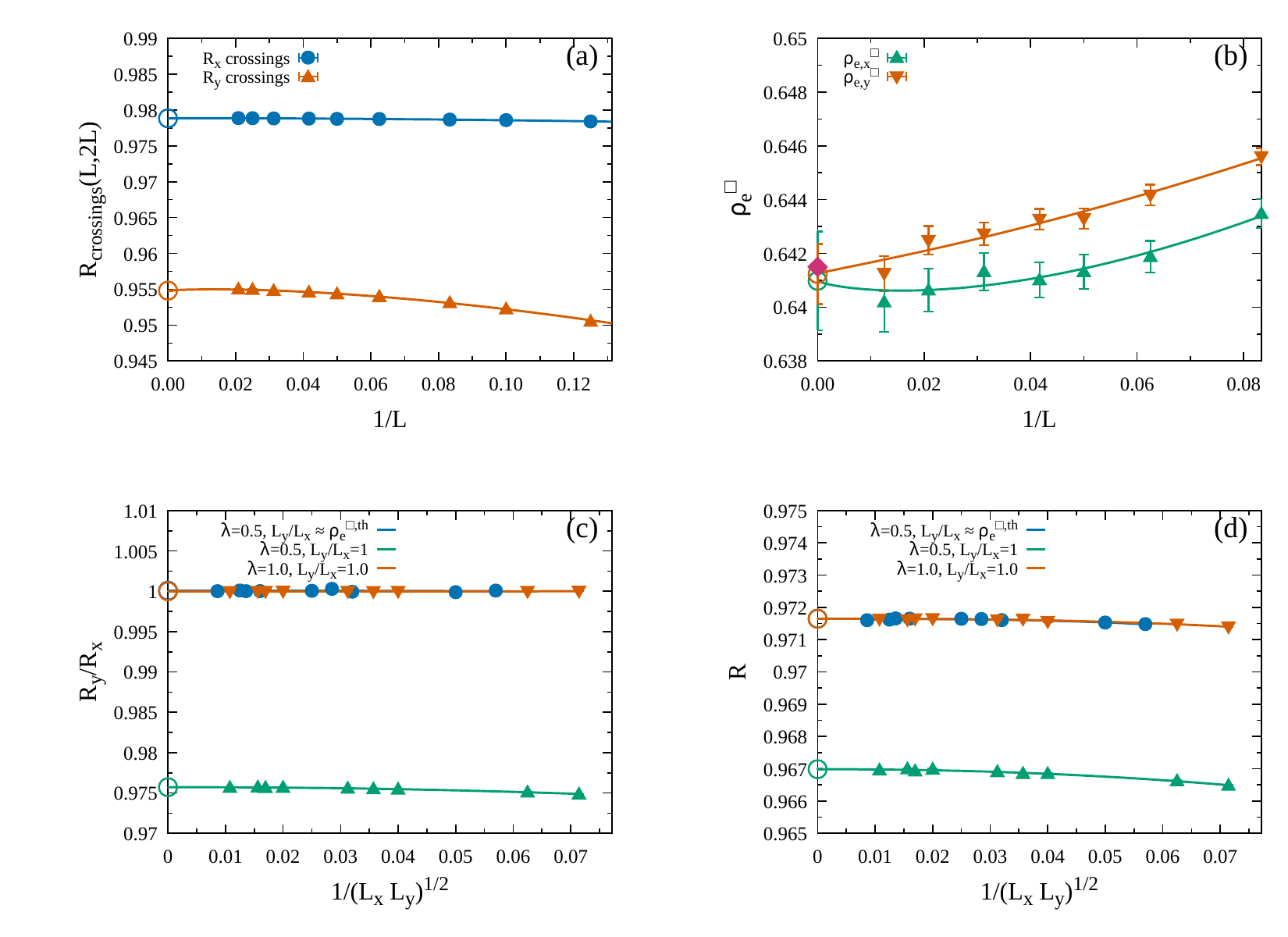}
\caption{Effective geometry and geometric compensation for $\lambda=0.5$. (a) Directional crossing values of $R_x$ and $R_y$ and their thermodynamic-limit extrapolations.
(b) Finite-size estimates of the effective aspect ratio $\rho_e^\square$ of a physically square anisotropic system inferred from the critical FK wrapping probabilities in the two directions.
The solid red symbol denotes the theoretical value $\rho_e^{\square,\mathrm{th}}$ obtained from the isoradial construction. (c) Ratio of the directional correlation ratios, $R_y/R_x$.
(d) Overall correlation ratio $R$. Panels (c) and (d) compare the anisotropic physical square at
$\lambda=0.5$, the geometrically compensated anisotropic rectangle at $\lambda=0.5$, and the isotropic square at $\lambda=1$, evaluated at their respective critical temperatures and plotted against $1/L_{\rm eff}$, where $L_{\rm eff}=(L_xL_y)^{1/2}$. For the compensated rectangles, the physical aspect ratio is chosen using the theoretical effective aspect ratio,
$L_y/L_x\simeq\rho_e^{\square,\mathrm{th}}$. Solid curves denote finite-size extrapolations, and open symbols at $1/L=0$ or $1/L_{\rm eff}=0$ denote the corresponding thermodynamic-limit values.}
\label{fig:geometry}
\end{figure}

To this end, we first consider the directional correlation ratios at
their crossings. For each size pair $(L,2L)$, the crossing value is defined by
$R_{\mu,\mathrm{cross}}(L,2L)=R_\mu[T_c^{(\mu)}(L),L]=R_\mu[T_c^{(\mu)}(L),2L]$, with $\mu=x,y$.
The data in Figure~\ref{fig:geometry}(a) are extrapolated using
\begin{equation}
R_{\mu,\mathrm{cross}}(L,2L)
=R_{\mu,c}+a_{R,\mu}L^{-\omega}+b_{R,\mu}L^{-2\omega},
\label{eq:directional-R-fit}
\end{equation}
where $\omega=4/5$ is the leading finite-size correction-to-scaling exponent of the two-dimensional three-state Potts universality class
\cite{PhysRevB.77.144410}, and the $L^{-2\omega}$ term accounts for a
higher-order finite-size correction. In contrast to the crossing temperatures and
correlation-length exponents, the two directional crossing values approach
distinct thermodynamic-limit values, $R_{x,c}=0.9791(2)$ and $R_{y,c}=0.95527(5)$. Thus, a common critical temperature and correlation-length exponent do not require the dimensionless directional critical ratios to become identical.

This behavior can be understood in terms of an anisotropic spatial metric. Close to criticality, the directional bulk correlation lengths may be written as
$\xi_\mu(t)\simeq\xi_{0,\mu}^{+}t^{-\nu}$, where $t=(T-T_c)/T_c\to0^+$ and $\mu=x,y$. The exponent $\nu$ is common to both directions, whereas the correlation-length amplitudes $\xi_{0,\mu}^{+}$ are direction dependent. Rescaling the coordinates according to
$x'=x/\xi_{0,x}^{+}$ and $y'=y/\xi_{0,y}^{+}$ removes the leading spatial anisotropy. Defining the effective aspect ratio as the ratio of the rescaled lengths in the $x$ and $y$ directions gives
\begin{equation}
\rho_e=\frac{L_x/\xi_{0,x}^{+}}{L_y/\xi_{0,y}^{+}}
=\frac{L_x}{L_y}\rho_e^\square,
\qquad
\rho_e^\square
=\frac{\xi_{0,y}^{+}}{\xi_{0,x}^{+}}.
\label{eq:rectangular-effective-aspect}
\end{equation}
Here, $\rho_e^\square$ denotes the effective aspect ratio of a physically square system.

A theoretical value of this effective aspect ratio can be obtained by specializing the isoradial construction for anisotropic Potts models of Ref.~\cite{HuZiffDeng2022} to the present square lattice. We introduce $v_\mu=e^{\beta_cJ_\mu}-1$ and $r_q=\pi^{-1}\arccos(\sqrt q/2)$. The corresponding isoradial angles satisfy
\begin{equation}
v_\mu
=\sqrt q\,\frac{\sin[r_q(\pi-\theta_\mu)]}{\sin(r_q\theta_\mu)}.
\label{eq:isoradial-angle}
\end{equation}
For the square lattice, the self-dual condition $v_xv_y=q$
corresponds to $\theta_x+\theta_y=\pi$. Since the principal axes of
the anisotropy coincide with the lattice axes, the transformation
preserves the rectangular geometry and introduces no boundary twist.
The anisotropy therefore reduces to a directional metric rescaling,
for which the effective aspect ratio of a physical square is
\begin{equation}
\rho_e^{\square,\mathrm{th}}=\frac{\sin(\theta_x/2)}{\sin(\theta_y/2)}.
\label{eq:rho-theory}
\end{equation}
For $q=3$, $J_x=1$, and $J_y=\lambda=0.5$, the exact critical inverse
temperature is $\beta_c^{\rm sd}=\ln 4$, yielding $v_x=3$, $v_y=1$,
and $r_q=1/6$. Equation~\eqref{eq:isoradial-angle} then determines
$\theta_x/\pi$ and $\theta_y/\pi$, and hence
$\rho_e^{\square,\mathrm{th}}\simeq0.64150030$.

Thus, a system that is square in the physical coordinates corresponds to a rectangle in the effective critical geometry. This provides a geometric explanation for why $R_x$ and $R_y$ can have distinct limiting critical values despite sharing the same $T_c$ and $\nu$. Since the principal axes of the anisotropy coincide with the lattice axes, the transformation preserves the rectangular geometry and the periodic boundary identifications without introducing a boundary twist. The resulting effective rectangle lacks direction-exchange
symmetry, so its directional critical ratios need not be equal. This geometric interpretation is consistent with the spatial rescaling framework of Ref.~\cite{HuZiffDeng2022}.

We independently determine the same effective aspect ratio from the critical Fortuin--Kasteleyn (FK) wrapping probabilities. The measured wrapping probabilities are matched to the corresponding isotropic three-state Potts reference functions using the same wrapping event,
aspect-ratio convention, and periodic boundary conditions. For $\lambda=0.5$, the theoretical effective aspect ratio satisfies $\rho_e^{\square,\mathrm{th}}<1$, corresponding to a shorter
correlation length in the $y$ direction than in the $x$ direction in the original anisotropic system. The $y$-direction wrapping probability $P_y$ is therefore expected to exhibit a stronger response to variations in the effective aspect ratio. Accordingly, the value
of $\rho_e^\square$ reported below is obtained by inverting the isotropic reference function for $P_y$, while the estimate inferred independently from $P_x$ serves as a consistency check. The resulting finite-size estimates $\rho_e^\square(L)$ are shown in Figure~\ref{fig:geometry}(b) and fitted to
\begin{equation}
\rho_e^\square(L)
=\rho_e^\square+a_\rho L^{-\omega}+b_\rho L^{-2\omega},
\label{eq:effective-aspect-fit}
\end{equation}
with $\omega=4/5$ and reference-length factors absorbed into the amplitudes. The numerical estimate is $\rho_e^\square=0.6413(5)$, in close agreement with the theoretical
prediction $\rho_e^{\square,\mathrm{th}}\simeq0.64150030$. The FK wrapping probabilities therefore provide an independent numerical determination of the effective aspect ratio.

A direct test of the effective-geometry interpretation is obtained by compensating the metric anisotropy through the physical shape. Requiring the rescaled domain to be square, $\rho_e=1$, gives
$L_y/L_x\simeq\rho_e^{\square,\mathrm{th}}\simeq0.64150030$ for $\lambda=0.5$. Thus, the theoretical effective aspect ratio directly specifies the physical aspect ratio required to compensate the anisotropic critical geometry. In the simulations, this target ratio can only be approximated because the side lengths must be integers, and the resulting small deviations are included in the finite-size extrapolations.

The effect of geometric compensation is seen most directly in the directional correlation ratios. For the isotropic model on a square geometry, the two spatial directions are equivalent, so that $R_x=R_y$ and hence $R_y/R_x=1$. For the anisotropic model on a physical square, this directional equivalence is lost. Although $R_x$ and $R_y$ are governed by the same bulk critical point and critical exponent, their critical values are different because the physical square corresponds to an anisotropic effective geometry. Consequently, $R_y/R_x$ approaches a value different from unity. Figure~\ref{fig:geometry}(c) shows that this directional asymmetry is removed by geometric compensation. When the physical aspect ratio is chosen according to
$L_y/L_x\simeq\rho_e^{\square,\mathrm{th}}$, the anisotropic system approaches $R_y/R_x=1$, in agreement with the isotropic-square reference. In other words, the unequal directional correlation ratios of the anisotropic physical square, $R_x\neq R_y$, become equivalent again, $R_x\simeq R_y$, after the effective geometry is compensated. This provides a direct demonstration that the directional difference is encoded in the critical metric rather than in distinct directional critical behavior.

Having established the recovery of directional equivalence, we next examine whether the same geometric compensation also restores the overall correlation ratio. Figure~\ref{fig:geometry}(d) compares the overall correlation ratio $R$ for an anisotropic physical square at $\lambda=0.5$, a geometrically compensated anisotropic rectangle at $\lambda=0.5$, and an isotropic square at $\lambda=1$, each evaluated at its respective critical temperature. The anisotropic physical square approaches a limiting value different from that of the isotropic square. After geometric compensation, however, the anisotropic rectangular system approaches the isotropic-square reference. The effect of geometric compensation therefore extends from the directional relation $R_x\simeq R_y$ to the overall correlation ratio, providing a further test of the effective-geometry description.

Taken together, Figures~\ref{fig:directional} and \ref{fig:geometry} establish a clear separation between universal critical scaling and nonuniversal spatial geometry. Spatial anisotropy produces pronounced directional finite-size effects and direction-dependent correlation-length amplitudes, while the two
directions nevertheless converge to the same critical temperature and correlation-length exponent. The remaining anisotropy is instead encoded in an effective spatial metric, quantitatively characterized by $\rho_e^\square$. The effective aspect ratio obtained independently from the critical FK wrapping probabilities, $\rho_e^\square=0.6413(5)$, agrees closely with the theoretical value $\rho_e^{\square,\mathrm{th}}\simeq0.64150030$ derived from the
isoradial construction. More importantly, using the theoretical value to compensate the physical aspect ratio restores the directional relation from $R_x\neq R_y$ toward $R_x\simeq R_y$ and simultaneously brings the overall correlation ratio toward the isotropic-square reference. This provides a direct geometric test of the effective-metric description. The correspondence concerns the critical scaling geometry and does not imply equality of the microscopic couplings or critical temperatures. Spatial anisotropy therefore primarily changes the critical metric and the associated finite-size geometry while leaving the bulk universality class of the two-dimensional three-state Potts model unchanged.

\section{Conclusions and Discussion}
\label{sec:conclusions}
We have investigated the critical behavior of the anisotropic two-dimensional ferromagnetic three-state Potts model on the square lattice using cluster Monte Carlo simulations, finite-size scaling, Fisher-zero analysis, and direction-resolved geometric observables. For coupling ratios $\lambda=0.5$, $0.75$, and $1$, the estimated critical temperatures remain close to the exact self-dual values, while the critical exponents remain close to the exact values of the
two-dimensional three-state Potts universality class. The Fisher-zero analysis provides an independent consistency check through estimates of the correlation-length and specific-heat exponents. These results show that spatial anisotropy changes nonuniversal quantities such as the critical temperature while leaving the bulk critical behavior consistent with the three-state Potts universality class over the range of anisotropy considered here.

The directional analysis reveals, however, that unchanged bulk universality does not imply isotropic finite-size critical behavior. For $\lambda=0.5$, the directional correlation ratios $R_x$ and $R_y$ exhibit different finite-size trajectories and approach distinct critical values on a physically square lattice, while their scaling is governed by a common critical temperature and correlation-length exponent. The anisotropy is therefore retained at criticality through direction-dependent length scales and dimensionless finite-size observables rather than through distinct directional correlation-length exponents. This behavior is consistent with the general picture in
which anisotropic critical systems can be related to isotropic reference systems through an appropriate rescaling of spatial lengths or system geometry \cite{BloteDeng2002,DengBlote2004,HuZiffDeng2022}.

A quantitative characterization of this rescaling is provided by the effective aspect ratio. From the critical FK wrapping probabilities, we obtain $\rho_e^\square=0.6413(5)$ for a physically square system at $\lambda=0.5$, in close agreement with the theoretical prediction $\rho_e^{\square,\mathrm{th}}\simeq0.64150030$ obtained from the isoradial construction \cite{HuZiffDeng2022}. Importantly, the wrapping probabilities provide an independent numerical determination of the effective aspect ratio, without using the directional correlation ratios. Using the theoretical value to compensate the physical aspect ratio restores the directional relation from
$R_x\neq R_y$ toward $R_x\simeq R_y$ and simultaneously brings the overall correlation ratio toward the isotropic-square reference. The agreement between the independently determined numerical effective aspect ratio and its theoretical prediction, together with the subsequent geometric-compensation test, provides direct numerical support for the effective-metric description of spatial anisotropy at criticality.

These results establish a useful separation between universal bulk
critical scaling and anisotropy-dependent finite-size geometry.
Spatial anisotropy can substantially modify directional dimensionless
observables without changing the underlying bulk universality class.
Unlike conventional scalar quantities such as the Binder cumulant,
which do not naturally resolve the two spatial directions, the
directional correlation ratios directly probe the anisotropic
finite-size geometry. The evolution from $R_x=R_y$ in the isotropic
square to $R_x\neq R_y$ in the anisotropic physical square, followed
by the recovery of $R_x\simeq R_y$ after geometric compensation,
makes the change in effective critical geometry directly visible.
The directional correlation ratios therefore provide a complementary
probe of anisotropic finite-size critical behavior
\cite{Tomita2002}.

It would be useful to test this effective-geometry description over a broader range of coupling anisotropies, particularly in regimes where finite-size corrections become stronger, and for other Potts models and lattice geometries. Such studies could clarify how generally direction-resolved correlation observables and wrapping probabilities can be used to characterize and compensate anisotropic critical geometry.
\begin{acknowledgments}
Y. L. acknowledges support from the National Natural Science Foundation of China under Grant Nos. 12305039 and 12574251, the Fundamental Research Funds for the Central Universities from the Beijing University of Posts and Telecommunications under Grant Nos. 2025JCTP08 and 2025AI4S01, as well as the Open Fund of the Key Laboratory of Multiscale Spin Physics (Ministry of Education), Beijing Normal University, under Grant No. SPIN2025K03. L. L. is supported by the National Natural Science Foundation of China under
Grant Nos. 12304171 and 12674288.
\end{acknowledgments}

\bibliography{fassaad}
\end{document}